\documentclass{elsart}
\usepackage{epsfig}
\usepackage{amssymb}

\begin{document}

\begin{frontmatter}

\title{Elliptic flow of thermal photons and dileptons}%\thanksref{DOE}}
\thanks[DOE]{Work supported by the U.S. Department of Energy (U.H.), 
by NSERC of Canada and by the Fonds Qu\'eb\'ecois de Recherche sur 
la Nature et les Technologies (C.G).}

\vspace*{-5mm}
\author[OSU]{{U.\,Heinz$\!$}\thanksref{me}},
\thanks[me]{Invited speaker and corresponding author. {\it Email:}  
\tt{heinz@mps.ohio-state.edu}}
\author[VECC]{R.\,Chatterjee$\!$},
\author[OSU]{E.\,Frodermann$\!$},
\author[McGill]{C.\,Gale$\!$},\,\,and\,\,%
\author[VECC]{D.\,K.\,Srivastava$\!$}
\address[OSU]{Department of Physics, The Ohio State University, 
Columbus, OH 43210, USA}
\address[VECC]{Variable Energy Cyclotron Centre, 1/AF Bidhan Nagar, 
Kolkata 700 064, India}  
\address[McGill]{Department of Physics, McGill University,
Montreal, Qu\'ebec H3A\,2T8, Canada}

\vspace*{-5mm}
\begin{abstract}
In this talk we describe the recently discovered rich phenomenology of 
elliptic flow of electromagnetic probes of the hot matter created in 
relativistic heavy-ion collisions. Using a hydrodynamic model for
the space-time dynamics of the collision fireball created in Au+Au 
collisions at RHIC, we compute the transverse momentum spectra and elliptic 
flow of thermal photons and dileptons. These observables are shown to
provide differential windows into various stages of the fireball 
expansion.
\end{abstract}

%\begin{keyword}
% keywords here, in the form: keyword \sep keyword
%Elliptic flow \sep hydrodynamics\sep thermal photons \sep thermal dileptons
% PACS codes here, in the form: \PACS code \sep code
%\PACS 25.75.-q \sep 12.38.Mh
%\end{keyword}
\end{frontmatter}

\parskip=3mm

\vspace*{-7mm}
%%%%%%%%%%%%%%%%%%% Main Text %%%%%%%%%%%%%%%%%%%%%%%%%%%%%%%%%%%%%%%%%%%%%%
\section{Introduction}
\label{sec1}
\vspace*{-5mm}
%%%%%%%%%%%%%%%%%%%%%%%%%%%%%%%%%%%%%%%%%%%%%%%%%%%%%%%%%%%%%%%%%%%%%%%%%%%%

The strong radial and elliptic flow of hadrons observed in 
relativistic heavy-ion collisions at RHIC have led to the important
conclusion that the quark-gluon plasma (QGP) created in these 
collisions acts like a strongly coupled plasma with almost perfect 
liquid behaviour. This conclusion is based on the successful prediction 
of the hadron momentum distributions, in particular of their 
anisotropies in non-central collisions, by dynamical calculations 
which treat the expanding QGP as an ideal fluid. While no other 
equally successful model exists, one has to remain conscious of the 
fact that the new ``perfect liquidity'' paradigm is based on a model 
back-extrapolation of the measured data to the early stages of the 
collision which are not directly accessible with hadronic observables. 
There are strong arguments that this back-extrapolation is fairly unique 
\cite{SQCD} and hence that the above-mentioned qualitative conclusion is  
robust. On a quantitative level, however, the extraction from experimental 
data of the (small) QGP viscosity is presently hampered not only by the 
unavailability of consistent hydrodynamic codes for viscous relativistic 
fluids, but even more by uncertainties about the hydrodynamic effects of 
changes in the equation of state of the QGP matter \cite{Pasi} and about 
details of the initial conditions at the beginning of the hydrodynamic 
expansion stage \cite{Hirano,Lappi}.

It would therefore be invaluable to have data on additional experimental 
observables which probe directly the earlier expansion stages and 
help to further constrain the space-time and momentum-space 
characteristics of the fireball expansion and the related model 
uncertainties. Electromagnetic probes, in particular direct photons 
and dileptons, provide such observables. Due to the weakness of the
electromagnetic interaction, real and virtual photons are not created
abundantly, making their measurement difficult, but once created they
escape the fireball without reinteraction, turning them into direct
probes of the conditions under which they were created. The shape 
of their transverse momentum and invariant mass spectra has long been 
advertised as a direct probe of the extremely hot temperatures during 
the earliest stages of the expanding fireball. We here study these 
spectra at the next finer level of detail, by analyzing their 
anisotropies, especially their elliptic flow, in non-central collisions.
Our goal is to use such measurements to illuminate with better resolution 
the early evolution of the fireball's spatial deformation through its 
imprint on the momentum anisotropies of the photons emitted during 
these early stages. Future studies will further complement the analysis 
presented here by directly measuring the space-time structure of the 
early collision fireball with two-photon correlations.

\vspace*{-6mm}
%%%%%%%%%%%%%%%%%%%%%%%%%%%%%%%%%%%%%%%%%%%%%%%%%%%%%%%%%%%%%%%%%%%%%%%%%%%%
\section{Spectra and elliptic flow}
\label{sec2}
%%%%%%%%%%%%%%%%%%%%%%%%%%%%%%%%%%%%%%%%%%%%%%%%%%%%%%%%%%%%%%%%%%%%%%%%%%%%
\vspace*{-5mm}

Both real and virtual photon (dilepton) momentum spectra can be 
written as 
\begin{equation}
\label{eq3}
  E\, dN/d^3p = \int\Bigl[(...)\exp(-p{\cdot}u(x)/T(x))\Bigr]\,d^4x\,,
\end{equation}
\vspace*{-4mm}
where the quantity inside the square brackets indicates the thermal 
emission rates from the QGP or hadronic matter. The photon 
4-momentum is parametrized by its rapidity $Y$, its transverse momentum 
$p_T{\,=\,}(p_x^2+p_y^2)^{1/2}$, and its azimuthal emission angle 
$\phi$ as $p^\mu{\,=\,}(M_T \cosh Y,p_T\cos\phi,p_T\sin\phi,p_T \sinh Y)$. 
For real photons $M_T=p_T$; for dileptons $M_T=[M^2+p_T^2]^{1/2}$ 
where $M$ is the invariant mass of the virtual photon and lepton pair. 
The spectral shape is dominated by the Boltzmann factor describing a 
flow-boosted thermal distribution. Assuming boost-invariant longitudinal
expansion and using standard proper time and space-time rapidity
coordinates, it is given by
\begin{equation}
\label{eq4}
 \frac{p \cdot u(x)}{T(x)} = \frac{\gamma_T(x)}{T(x)} 
 \Bigl[M_T \cosh (Y{-}\eta)- p_T v_T(x)\cos(\phi{-}\phi_v(x))\Bigr]\,,
\end{equation}
\vspace*{-4mm}
where $v_T$ (with $\gamma_T{\,=\,}(1{-}v_T^2)^{-1/2}$) is the magnitude
and $\phi_v{\,=\,}\tan^{-1}(v_y/v_x)$ the azimuthal angle of the tranverse 
flow velocity. 

The azimuthal anisotropy ($\phi$-dependence) of the spectrum is controlled 
by an interplay between the collective flow anisotropy and the geometric 
deformation of the temperature field $T(x,y,\tau)$ at non-zero impact 
parameter. In the present work both are given by a hydrodynamical 
calculation, using the boost invariant hydrodynamic code AZHYDRO 
\cite{AZHYDRO} with standard \cite{AZHYDRO} initial conditions for 
Au+Au collisions at $\sqrt{s}{\,=\,}200\,A$\,GeV. The only adjustment we
make \cite{PRL} is to extrapolate the initial entropy density from the 
usual initial time $\tau_0{\,=\,}0.6$\,fm/$c$ to a 3 times smaller value 
$\tau_0{\,=\,}0.2$\,fm/$c$, assuming 1-dimensional boost-invariant 
expansion between these times, in order to account for at least a fraction 
of the pre-equilibrium photon production at very early times \cite{BMS04}.
Its contribution to the photon spectrum is important at large $p_T$,
and it will suppress its anisotropy there because very little 
transverse flow develops before 0.6\,fm/$c$. 

We concentrate on photons and dileptons emitted at midrapidity $Y{\,=\,}0$
so that the spectrum has only even azimuthal Fourier components $v_n$.
The elliptic flow $v_2$ is computed as the angular average of $\cos(2\phi)$
with the spectrum (\ref{eq3}) as weight function: 
$v_2{\,=\,}\langle\cos(2\phi)\rangle$.

\vspace*{-5mm}
%%%%%%%%%%%%%%%%%%%%%%%%%%%%%%%%%%%%%%%%%%%%%%%%%%%%%%%%%%%%%%%%%%%%%%%%%%%%
\section{Thermal photons \cite{PRL}}
\label{sec3}
%%%%%%%%%%%%%%%%%%%%%%%%%%%%%%%%%%%%%%%%%%%%%%%%%%%%%%%%%%%%%%%%%%%%%%%%%%%%
\vspace*{-5mm}

\begin{figure}[h]
\begin{minipage}{6.8cm}
\begin{center}
\epsfig{file=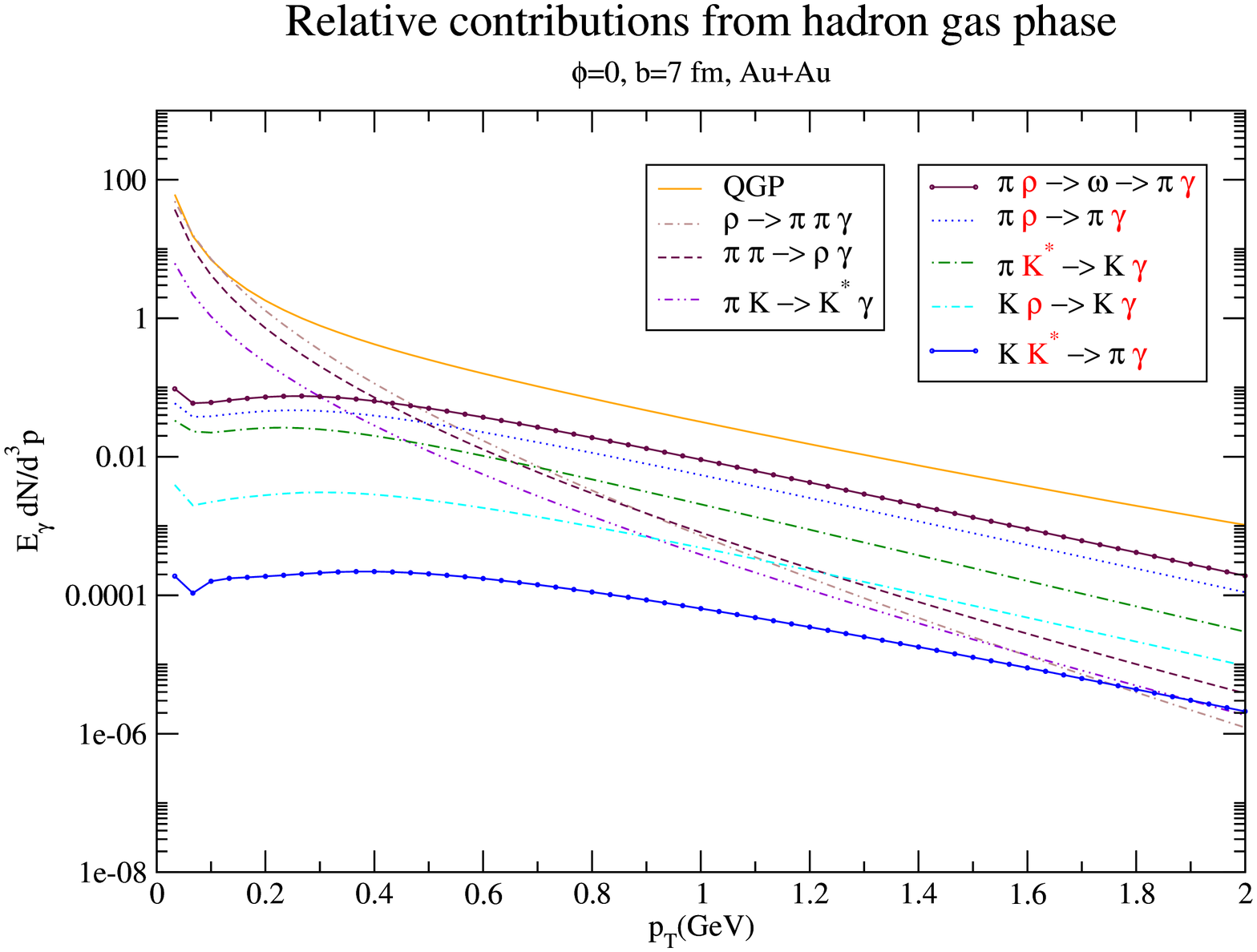,bb=61 0 590 710,height=6.8cm,angle=270,clip=}
\end{center}
\caption{Contributions to the photon $p_T$-spectrum at 
$\phi{\,=\,}0$ from the QGP and from various hadronic rescattering 
channels during the late hadron gas phase.}
\label{F1}
\end{minipage}
\hspace*{1mm}
\begin{minipage}{6.75cm}
\begin{center}
\vspace*{3mm}
\epsfig{file=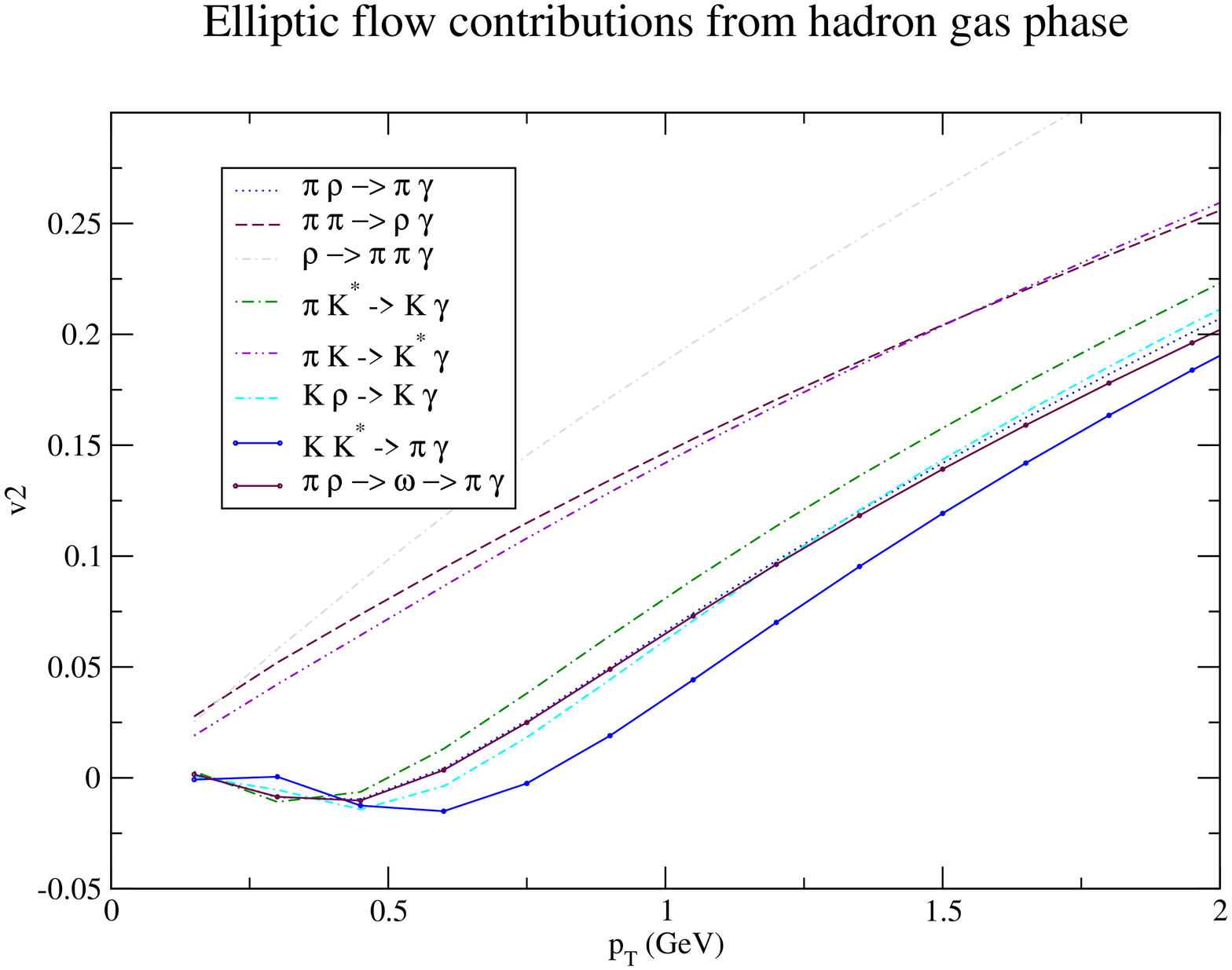,bb=90 30 590 710,height=6.75cm,width=4.82cm,angle=270,clip=}
\end{center}
\caption{Elliptic flows $v_2(p_T)$ associated with the individual hadronic
photon contributions shown in Fig.~\ref{F1} (Au+Au at $b{\,=\,}7$\,fm in 
both figures).}
\label{F2}
\end{minipage}
\end{figure}

For the photon emission rate from the QGP phase we use the complete 
leading-order expression from Ref.~\cite{guy}, while the latest 
results in Ref.~\cite{simon} are used for photon radiation from the 
hadron gas phase. The hadronic emission rate sums over a large number
of hadronic rescattering channels. Figure~\ref{F1} shows that these 
cluster into two different classes: at low $p_T$, photon production 
is dominated by {\em radiation} accompanying vector meson production 
from or decay into pions, whereas at large $p_T$ collision induced 
{\it conversion} of vector mesons into photons dominates. Correspondingly 
%
%%%%%%%%%%%%%%%%%%%%%%%%%%% Figs. 1 & 2 %%%%%%%%%%%%%%%%%%%%%%%%%%%%%%%%%%%%
%%%%%%%%%%%%%%%%%%%%%%%%%%%%%%%%%%%%%%%%%%%%%%%%%%%%%%%%%%%%%%%%%%%%%%%%%%%%
%
the emitted photons trace the {\em pion} momentum distribution at low $p_T$ 
and the {\em vector meson} momentum distribution at higher $p_T$. The latter 
is flattened by the effects of radial collective flow on the heavy vector 
mesons, exhibiting, in fact, a weak ``blast wave peak'' around 
$p_T{\,=\,}0.4-0.5$\,GeV/$c$. This also explains why in Figure~\ref{F2} the 
photons from $\pi\pi{\,\to\,}\rho\gamma$ etc. scattering exhibit the standard 
almost linear rise of the elliptic flow $v_2$ with $p_T$, well known from 
the pions themselves, whereas the vector meson conversion photons carry much 
less elliptic flow, even contributing negatively at low $p_T$ below the 
``blast wave peak'' in their $p_T$ spectrum.

%
%%%%%%%%%%%%%%%%%%%%%%%%%%% Fig. 3 %%%%%%%%%%%%%%%%%%%%%%%%%%%%%%%%%%%%%%%%%
\begin{figure}[h]
\begin{center}
\epsfig{file=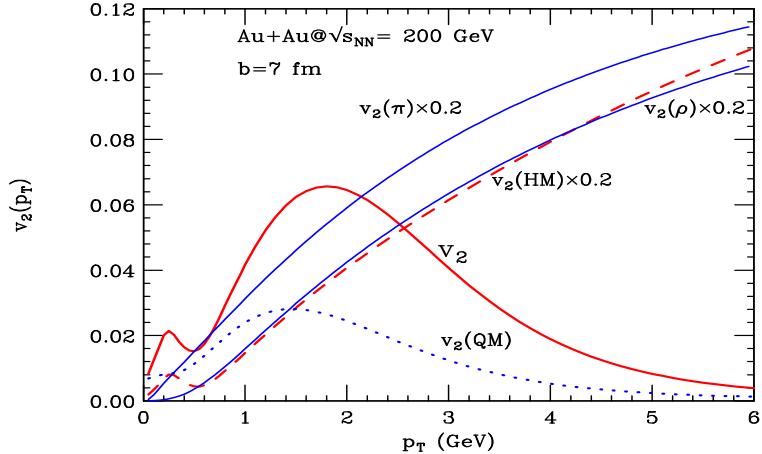,width=10cm,height=6cm}
\end{center}
\caption{Total photon elliptic flow $v_2(p_T)$, as well as hadronic
and quark matter contributions for comparison. (Figure taken from
\cite{PRL}.)} 
\label{F3}
\end{figure}
%%%%%%%%%%%%%%%%%%%%%%%%%%%%%%%%%%%%%%%%%%%%%%%%%%%%%%%%%%%%%%%%%%%%%%%%%%%%
%

Figure~\ref{F3} shows the total thermal photon elliptic flow (solid 
red line) and compares it with the elliptic flow of the early quark 
matter photons (dotted blue line) and of the late hadronic matter
photons (dashed red line). Consistent with the spectra shown in 
Fig.~\ref{F1}, the hadron gas photons are seen to track the elliptic
flow of pions at low $p_T$ and that of $\rho$ mesons at higher $p_T$ 
(both shown by solid blue lines in Fig.~\ref{F3} and exhibiting 
the almost linear rise of $v_2$ with $p_T$ that is characteristic of 
hadrons in the hydrodynamic model \cite{QGP3}). The cross-over 
between these two patterns causes a distinctive peak-valley structure 
in $v_2$ around $p_T{\,=\,}0.4-0.5$\,GeV/$c$. Since the hydrodynamic 
model is known to work quantitatively very well in this transverse 
momentum range \cite{QGP3}, we expect this structure to be robust 
\cite{fn1}. 

In contrast to the hadronic photons, the quark matter photons show an
elliptic flow which decreases at high $p_T$. This reflects the fact 
that they track quark momenta, and that quark flow is small at early
times when the high-$p_T$ photons are emitted. As one goes down in $p_T$
one probes later emission times and sees an increasing $v_2$ of the quark 
matter photons, reflecting the buildup of elliptic flow in the quark fluid.
The behavior of the total photon elliptic flow, finally, can be understood
by realizing that hadronic photons dominate the total photon spectrum
only at low $p_T$ while quark matter radiation begins to take over around
$p_T{\,\simeq\,}0.4$\,GeV/$c$, completely outshining the hadron gas for
$p_T{\,>\,}1-2$\,GeV/$c$. This cuts off the linear rise of the hadronic
photon $v_2$, and the total photon elliptic flow at high transverse 
momenta thus reflects the small elliptic flow during the very early
collision stages.

One should remember, however, that ideal fluid dynamics gradually breaks 
down at higher $p_T$. Data suggest \cite{RANP04} that near the 
hadronization point quark elliptic flow begins to be seriously affected 
by viscous effects for $p_T{\,>\,}1$\,GeV/$c$, and this threshold may be
even lower at earlier times when the longitudinal expansion rate is higher
and shear viscous effects are larger. For $p_T{\,>\,}1$\,GeV/$c$ our 
hydrodynamic prediction of photon elliptic flow must thus be regarded as 
an upper limit, and its already small values at large $p_T$ will be 
further reduced by viscous corrections and prompt photon contributions 
\cite{fms_phot,gale}.

\vspace*{-5mm}
%%%%%%%%%%%%%%%%%%%%%%%%%%%%%%%%%%%%%%%%%%%%%%%%%%%%%%%%%%%%%%%%%%%%%%%%%%%%
\section{Thermal dileptons \cite{inprep}}
\label{sec4}
%%%%%%%%%%%%%%%%%%%%%%%%%%%%%%%%%%%%%%%%%%%%%%%%%%%%%%%%%%%%%%%%%%%%%%%%%%%%
\vspace*{-5mm}

\begin{figure}[h]
\begin{minipage}{6.8cm}
\begin{center}
\epsfig{file=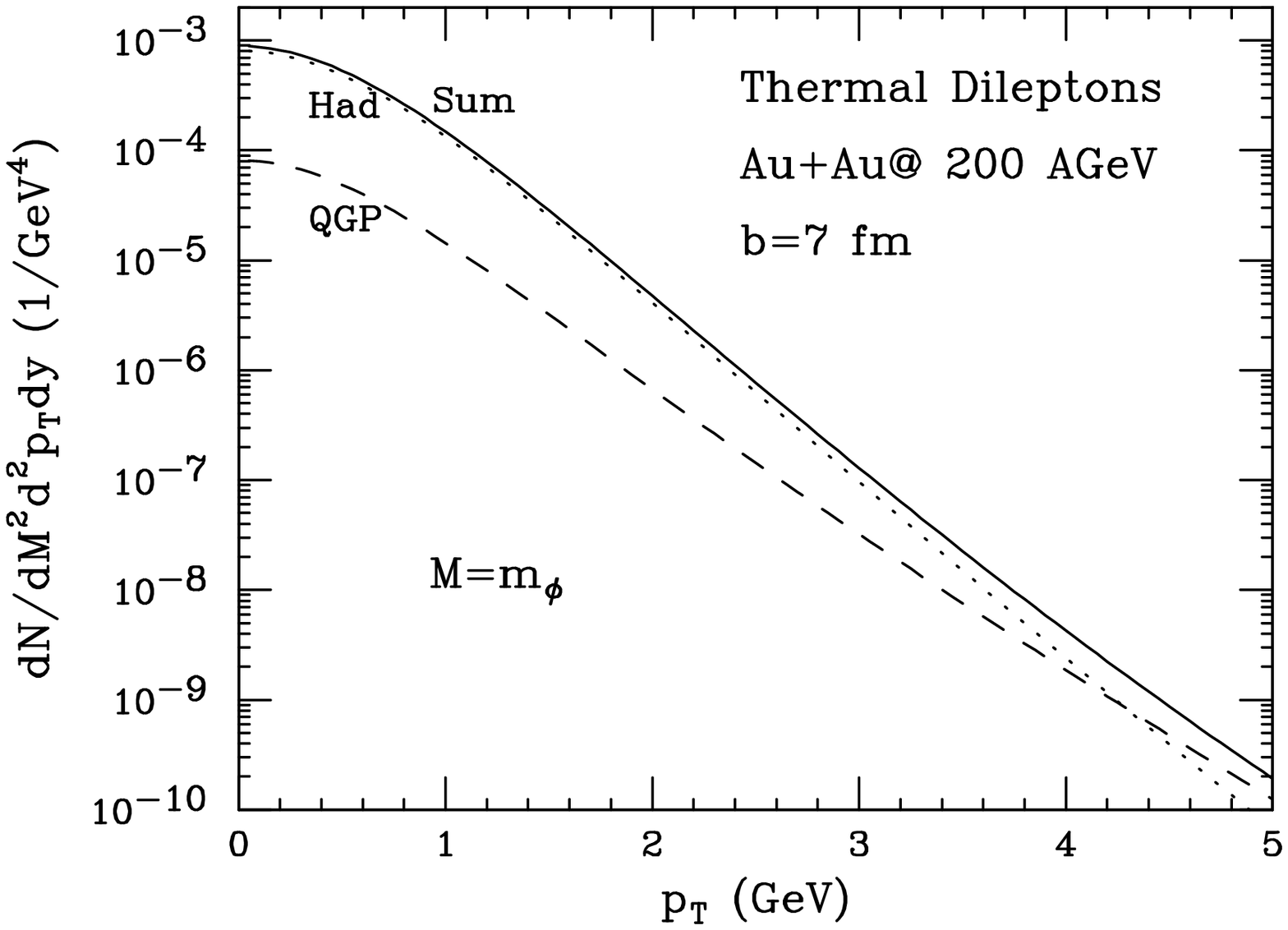,width=6.8cm}
\end{center}
\caption{The $p_T$-spectrum of thermal dileptons with mass $M{\,=\,}m_\phi$,
showing the total spectrum as well as separate contributions from quark
and hadronic matter.} 
\label{F4}
\end{minipage}
\hspace*{1mm}
\begin{minipage}{6.75cm}
\begin{center}
\epsfig{file=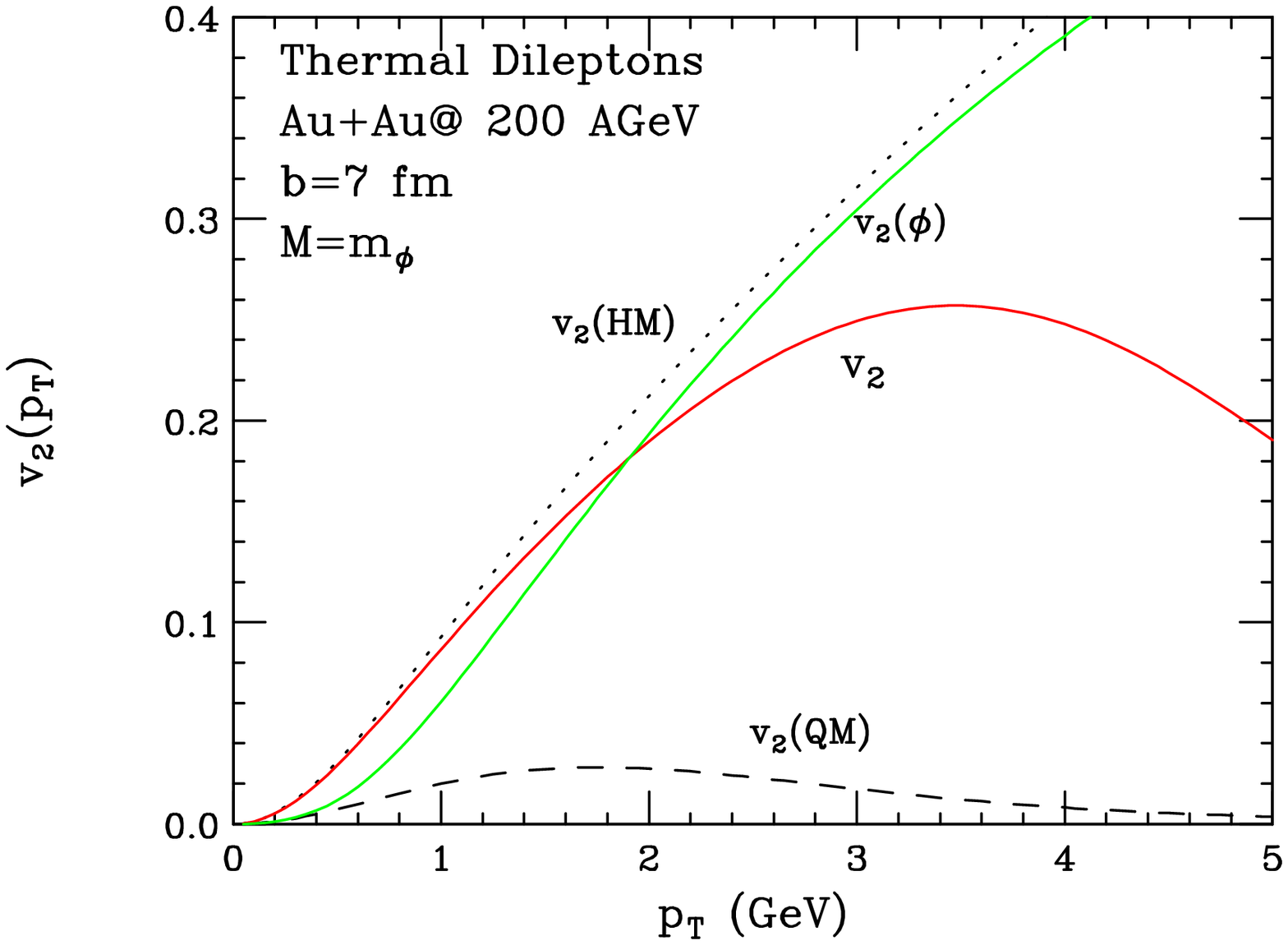,width=6.75cm}
\end{center}
\caption{$v2(p_T)$ of thermal dileptons with $M=m_\phi$ (of quark and 
hadronic matter dileptons and of their sum), as well as of $\phi$ 
mesons emitted at thermal freeze-out.}
\label{F5}
\end{minipage}
\end{figure}

With dileptons (virtual photons) we can probe the elliptic flow as
a function of an additional variable, their invariant mass 
$M{\,=\,}M_{\ell\bar\ell}{\,=\,}M_{\gamma^*}$: 
\begin{equation}
\label{eq5}
  v_2(M,p_T;b)  = 
  \frac{\int d\phi\,\cos(2\phi)\,
        \frac{dN_{\ell\bar\ell}(b)}{dM^2\,dY\,p_Tdp_T\,d\phi}}
       {\int d\phi\,
        \frac{dN_{\ell\bar\ell}(b)}{dM^2\,dY\,p_Tdp_T\,d\phi}}.
\end{equation}
\vspace*{-3mm}
In Figures~\ref{F4} and \ref{F5} we show $p_T$-spectra and elliptic
flow of thermal dileptons with invariant mass $M{\,=\,}m_\phi$. We see 
in Fig.~\ref{F4} that for this value of $M$ the $p_T$-spectrum
is completely dominated by virtual photon emission from the hadronic 
phase, all the way up to $p_T{\,=\,}4$\,GeV/$c$. The elliptic flow
%
%%%%%%%%%%%%%%%%%%%%%%%%%%% Figs. 4 & 5 %%%%%%%%%%%%%%%%%%%%%%%%%%%%%%%%%%%%
%%%%%%%%%%%%%%%%%%%%%%%%%%%%%%%%%%%%%%%%%%%%%%%%%%%%%%%%%%%%%%%%%%%%%%%%%%%%
%
of the dileptons emitted from the hadronic phase closely tracks that
of $\phi$-mesons at thermal freeze-out, as seen in Fig.~\ref{F5}.
The elliptic flow of hadronic dileptons with $M{\,=\,}m_\phi$ 
(blue dotted line) is slightly larger than that $\phi$ mesons (green
solid line) since radial flow (which suppresses $v_2$) continues to 
build up during the hadronic phase and the hadronic dileptons are 
on average emitted somewhat earlier than the $\phi$ mesons \cite{fn2}. 

We have checked \cite{inprep} that the calculated $\phi$ meson 
$p_T$-spectrum at thermal freeze-out agrees with the measured 
spectrum of $\phi$ mesons reconstructed from $K^+K^-$ decays by 
the PHENIX Collaboration \cite{phenix_phi}.
The elliptic flow of quark matter dileptons with $M{\,=\,}m_\phi$ 
(black dashed line in Fig.~\ref{F5}) is much smaller and shows the 
same characteristic decrease at large $p_T$ as the elliptic flow of 
thermal photons in Fig.~\ref{F3}, reflecting their early emission when 
the flow anisotropy of the quark fluid was still small. Due to the dominance
of hadronic dileptons at this invariant mass, this decrease of $v_2$
at large $p_T$ is seen in the overall elliptic flow only at very
large $p_T{\,>\,}4$\,GeV/$c$; below $p_T{\,=\,}2$\,GeV/$c$, the total
dilepton elliptic flow follows almost perfectly the hadronic $v_2$. 
\vspace*{5mm}

%
%%%%%%%%%%%%%%%%%%%%%%%%%%% Figs. 6 & 7 %%%%%%%%%%%%%%%%%%%%%%%%%%%%%%%%%%%%
\begin{figure}[h]
\begin{minipage}{6.8cm}
\begin{center}
\epsfig{file=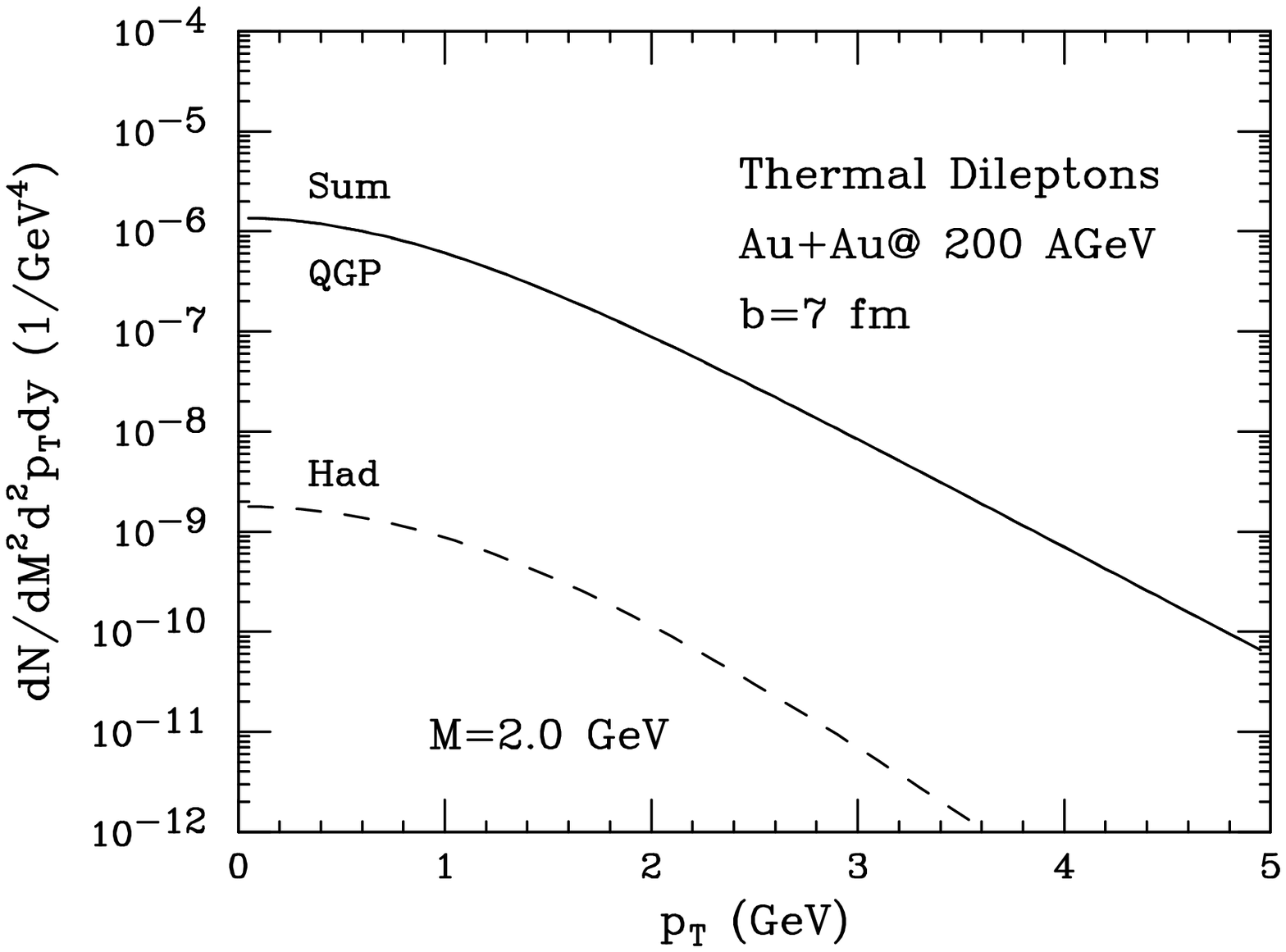,width=6.8cm}
\end{center}
\caption{Same as Fig~\ref{F4}, but for $M{\,=\,}2$\,GeV.}
\label{F6}
\end{minipage}
\hspace*{1mm}
\begin{minipage}{6.75cm}
\begin{center}
\epsfig{file=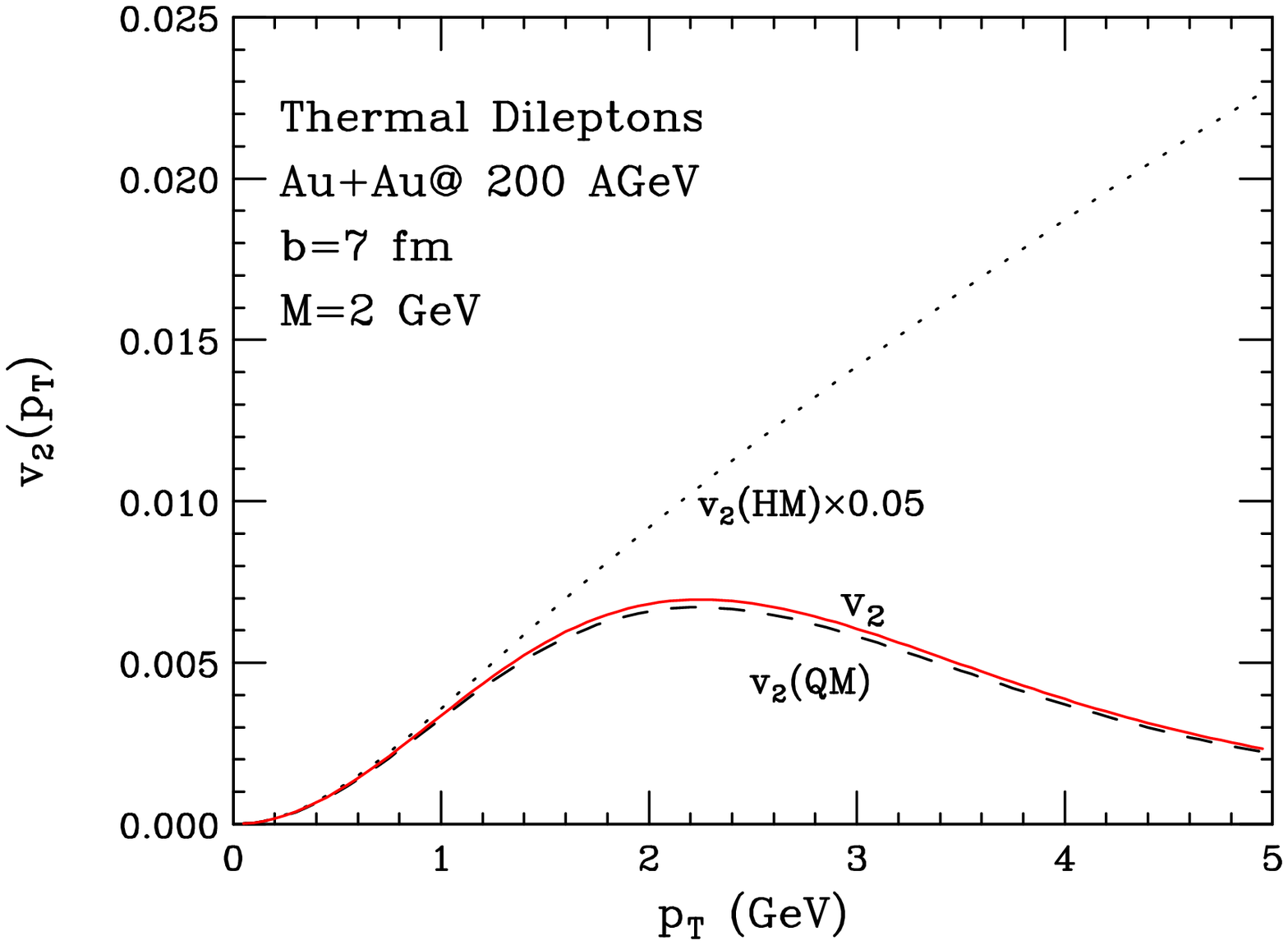,width=6.75cm}
\end{center}
\caption{Same as Fig~\ref{F5}, but for $M{\,=\,}2$\,GeV.}
\label{F7}
\end{minipage}
\end{figure}
%%%%%%%%%%%%%%%%%%%%%%%%%%%%%%%%%%%%%%%%%%%%%%%%%%%%%%%%%%%%%%%%%%%%%%%%%%%%
%
Qualitatively the same pattern is observed for dileptons with invariant
mass $M{\,=\,}m_\rho$ when compared with $\rho$ mesons emitted at 
thermal freeze-out \cite{inprep}. But things are different for dileptons 
with an invariant mass of $M{\,=\,}2$\,GeV (Figures~\ref{F6} and
\ref{F7}). Now the QGP dileptons outshine those from the hadron gas for
all $p_T$ by three orders of magnitude or more. While the elliptic flow
of the hadronic dileptons still shows the hydrodynamic almost linear
rise with $p_T$ (dotted line in Fig.~\ref{F7}), they are completely 
buried underneath the quark matter dileptons, and the total thermal
dilepton spectrum at $M{\,=\,}2$\,GeV thus exhibits clearly the rise 
and fall of the elliptic flow with increasing $p_T$ that is characteristic
of emission from the early QGP phase.

\begin{figure}[ht]
\begin{minipage}{6.8cm}
\begin{center}
\epsfig{file=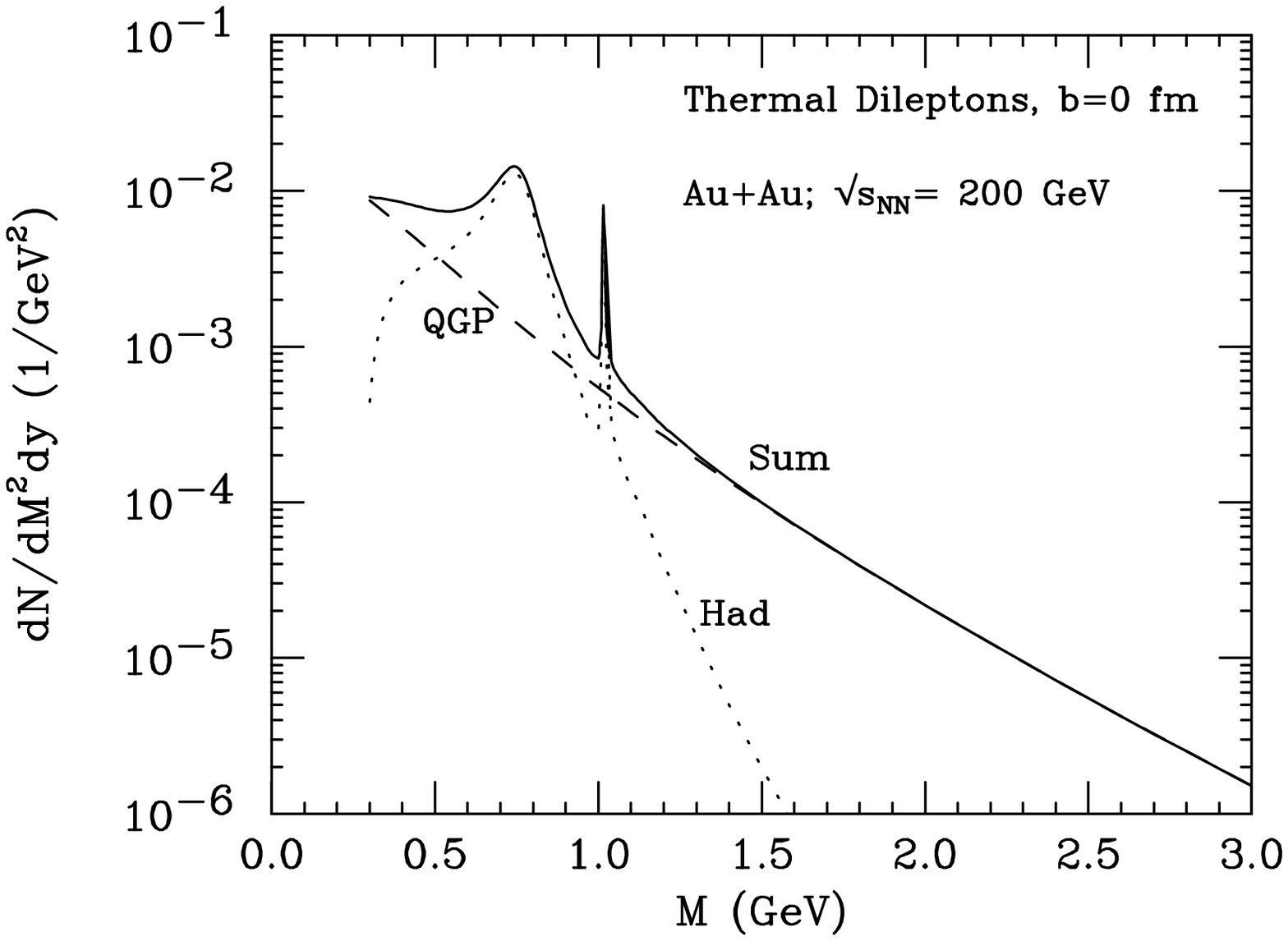,width=6.8cm}
\end{center}
\caption{QGP and hadron gas contributions to the dilepton invariant mass
spectrum for 200\,$A$\,GeV Au+Au collisions at $b{\,=\,}7$\,fm. The solid
line is the total dilepton mass spectrum.}
\label{F8}
\vspace*{1cm}
\end{minipage}
\hspace*{1mm}
\begin{minipage}{6.75cm}
\begin{center}
\epsfig{file=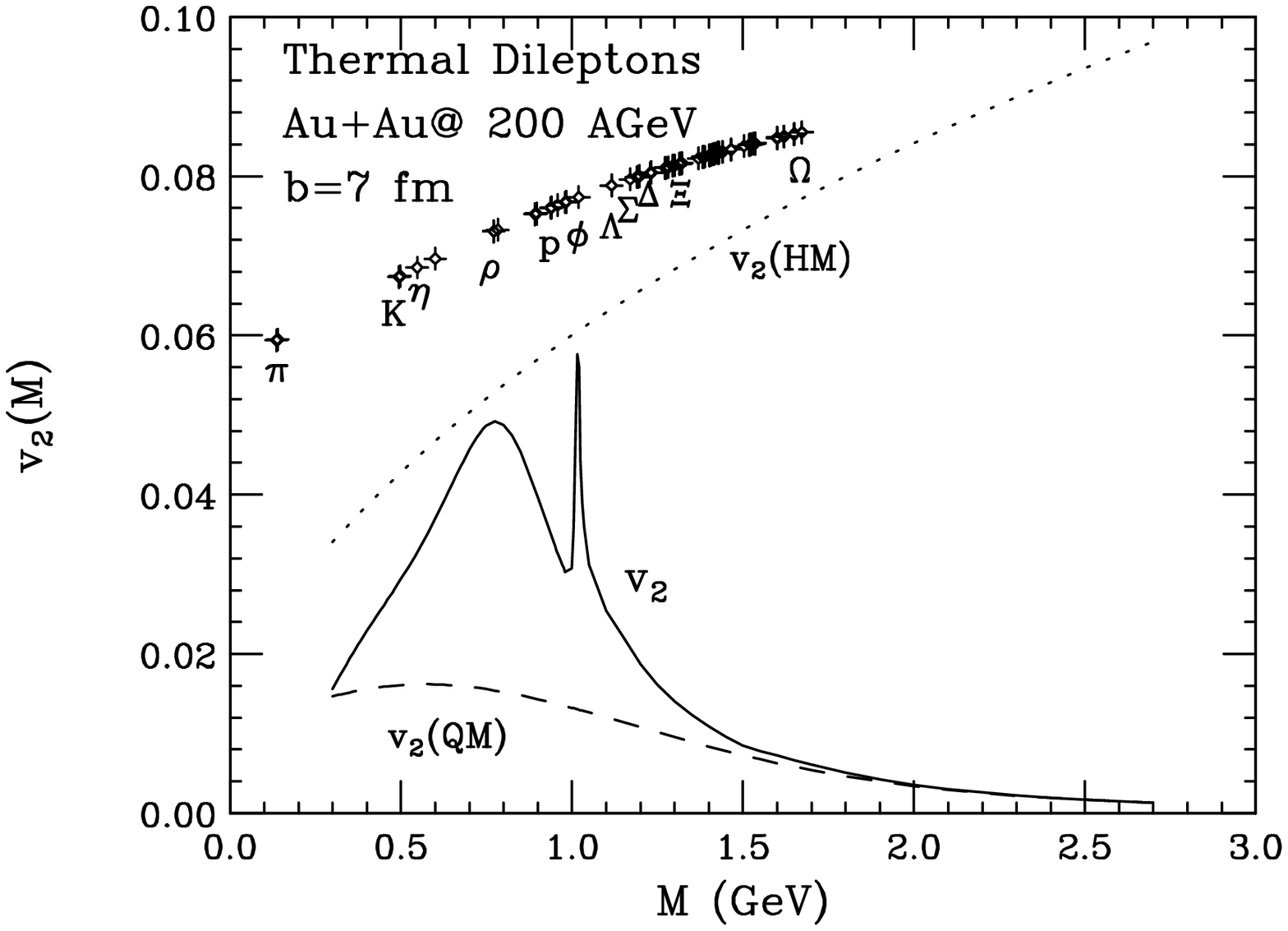,width=6.75cm}
\end{center}
\caption{The invariant mass dependence of the total $p_T$-integrated 
dilepton elliptic flow, as well as that of its hadron and quark matter 
contributions. Shown for comparison are also the $p_T$-integrated
elliptic flows for a number of hadrons emitted at thermal freeze-out.} 
\label{F9}
\end{minipage}
\end{figure}

The story is further clarified by studying the invariant mass dependence
of the $p_T$-integrated elliptic spectrum (the ``dilepton mass spectrum'',
Figure~\ref{F8}) and of the corresponding $p_T$-integrated elliptic
flow $v_2(M)$ (Figure~\ref{F9}). We note that these spectra are preliminary
and do not properly account for the $\omega$ meson; complete mass spectra 
of the dilepton elliptic flow which include the $\omega$ contribution
as well as dileptons from vector meson decays after thermal freeze-out
will be presented in \cite{inprep}. Figure~\ref{F8} shows that near the
vector mesons peaks ($\rho$, $\omega$, $\phi$), dilepton emission from 
the late hadronic stage dominates by at least an order of magnitude over
QGP radiation. Correspondingly, the total dilepton elliptic flow approaches 
in these mass regions the value of the hadronic matter dileptons (dotted 
line in Fig.~\ref{F9}). If one adds the post-freeze-out vector meson
decays, the total elliptic flow around $M{\,=\,}m_\omega$ and 
%
%%%%%%%%%%%%%%%%%%%%%%%%%%% Figs. 8 & 9 %%%%%%%%%%%%%%%%%%%%%%%%%%%%%%%%%%%%
%%%%%%%%%%%%%%%%%%%%%%%%%%%%%%%%%%%%%%%%%%%%%%%%%%%%%%%%%%%%%%%%%%%%%%%%%%%%
%
$M{\,=\,}m_\omega$ even approaches the elliptic flow of the corresponding
hadrons, $\omega$ and $\phi$, emitted at kinetic freeze-out. This last 
observation is explained by the relatively long lifetimes of the $\omega$
and $\phi$ mesons (23 and 45\,fm/c, respectively), which significantly
exceeds the duration of the hadron gas phase (${\sim\,}7$\,fm/$c$) 
such that on the resonance peaks dilepton emission from the hadron phase 
is overwhelmed by post-freeze-out dilepton decays of these resonances.

Note that the elliptic flow of the hadronic dileptons (dotted line in 
Fig.~\ref{F9}) remains significantly below that of the hadrons emitted
at thermal freeze-out at $T_\mathrm{dec}{\,=\,}130$\,MeV (crossed circles
in Fig.~\ref{F9}). In view of the similarity of the $p_T$-dependences
of their elliptic flows (see, e.g., the blue dotted and solid green lines
in Fig.~\ref{F5} this looks surprising. The puzzle is resolved by noting
that, at a fixed invariant mass, the hadronic dilepton $p_T$-spectrum is 
steeper than the thermal freeze-out hadron spectrum for a hadron with the 
same mass \cite{inprep}. The difference is likely due to additional radial 
flow built up between the average time of hadronic photon emission and 
final hadron freeze-out.

\vspace*{-5mm}
%%%%%%%%%%%%%%%%%%%%%%%%%%%%%%%%%%%%%%%%%%%%%%%%%%%%%%%%%%%%%%%%%%%%%%%%%%%%
\section{Conclusions}
\label{sec5}
%%%%%%%%%%%%%%%%%%%%%%%%%%%%%%%%%%%%%%%%%%%%%%%%%%%%%%%%%%%%%%%%%%%%%%%%%%%%
\vspace*{-5mm}

Elliptic flow of thermal photons and dileptons was shown to be a versatile
and potentially powerful probe of the fireball dynamics at RHIC and LHC,
complementary to the already well-studied flow anisotropies in the hadronic
sector. Contrary to hadron elliptic flow, which in a hydrodynamic picture
rises monotonically with increasing $p_T$ and in real life, due to viscous
effects, saturates at high $p_T$, photon and dilepton elliptic flow decrease
to zero at large $p_T$ and large dilepton mass, reflecting direct emission 
from the early QGP at a stage when flow anisotropies have not yet had time to
grow strong. $v_2^\gamma(p_T)$ and $v_2^{\ell\bar\ell}(M)$ exhibit rich
structures which reflect the interplay of different emission processes,
opening a window on detailed and differential information from a variety
of different stages of the fireball expansion. The elliptic flow of photons
and dileptons emitted from the late hadronic stage was seen to track the
$v_2$ of the emitting hadrons, which suggests the possibility of subtracting
the hadronic photon contributions from the total (virtual) photon signal
in order to isolate and study in greater detail the elliptic flow of early
QGP photons.

Obviously, the measurements will be difficult and the theoretical treatment 
can use further refinement, but this first glimpse suggests that photon and 
dilepton elliptic flow have the potential of turning into profitable gold 
mines. 

\vspace*{-5mm}
%%%%%%%%%%%%%%%%%%%%%%%%%%% References %%%%%%%%%%%%%%%%%%%%%%%%%%%%%%%%%%%


\begin{thebibliography}{00}
\vspace*{-5mm}
\itemsep=-1pt

\bibitem{SQCD}
  U.~Heinz and P.~F.~Kolb,
  %``Early thermalization at RHIC,''
  Nucl.\ Phys.\ A {\bf 702}, 269 (2002).
  % [arXiv:hep-ph/0111075].
  %%CITATION = HEP-PH 0111075;%%

\bibitem{Pasi}
  P.~Huovinen,
  %``Anisotropy of flow and the order of phase transition in relativistic  
  %  heavy ion collisions,''
  Nucl.\ Phys.\ A {\bf 761}, 296 (2005).
  % [arXiv:nucl-th/0505036].
  %%CITATION = NUCL-TH 0505036;%%

\bibitem{Hirano}
  T.~Hirano, U.~Heinz, D.~Kharzeev, R.~Lacey and Y.~Nara,
  % ``Hadronic dissipative effects on elliptic flow in ultrarelativistic
  %heavy-ion collisions,''
  Phys.\ Lett.\ B {\bf 636}, 299 (2006).
  % [arXiv:nucl-th/0511046].
  %%CITATION = NUCL-TH 0511046;%%

\bibitem{Lappi}
  T.~Lappi and R.~Venugopalan,
  % ``Universality of the saturation scale and the initial eccentricity 
  % in heavy ion collisions,''
  arXiv:nucl-th/0609021.
  %%CITATION = NUCL-TH 0609021;%%

\bibitem{AZHYDRO}
  The code can be downloaded from URL 
  http://nt3.phys.columbia.edu/people/ molnard/OSCAR/. 
  See also P.~F.~Kolb, J.~Sollfrank and U.~Heinz,
  %``Anisotropic transverse flow and the quark-hadron phase transition,''
  Phys.\ Rev.\ C {\bf 62}, 054909 (2000), and 
  P. F. Kolb and R. Rapp, Phys. Rev. C {\bf 67}, 044903 (2003). 

\bibitem{PRL}
  R.~Chatterjee, E.~S.~Frodermann, U.~Heinz and D.~K.~Srivastava,
  %``Elliptic flow of thermal photons in relativistic nuclear collisions,''
  Phys.\ Rev.\ Lett.\  {\bf 96}, 202302 (2006).
  % [arXiv:nucl-th/0511079].
  %%CITATION = NUCL-TH 0511079;%%

\bibitem{BMS04}
  S.~A.~Bass, B.~M\"uller and D.~K.~Srivastava,
  %``Intensity interferometry of direct photons emitted in Au + Au
  %collisions,''
  Phys.\ Rev.\ Lett.\ {\bf 93}, 162301 (2004).
  % [arXiv:nucl-th/0404050].
  %%CITATION = NUCL-TH 0404050;%%

\bibitem{guy} 
  P.~Arnold, G.~D.~Moore, and L.~G.~Yaffe, JHEP {\bf 0112}, 009 (2001).

\bibitem{simon} 
  S.~Turbide, R.~Rapp, and C.~Gale, 
  Phys. \ Rev. \ C {\bf 69}, 014903 (2004).

\bibitem{QGP3}
  P.~F.~Kolb and U.~Heinz, in {\it Quark-Gluon Plasma 3}, edited by 
  R.C. Hwa and X.-N. Wang (World Scientific, Singapore, 2004), p.~634 
  [nucl-th/0305084].

\bibitem{fn1} 
Some details may still change when hadron chemical 
freeze-out directly after hadronization near $T_\mathrm{c}{\,=\,}170$\,MeV 
is properly taken into account, by assigning all hadrons appropriate 
$T$-dependent chemical potentials (see discussion in \cite{AZHYDRO}) and 
folding these also into the photon emission rates. This was not 
done in \cite{PRL}.

\bibitem{RANP04}
  U.~Heinz,
  %``Thermalization at RHIC,''
  AIP Conf.\ Proc.\  {\bf 739}, 163 (2005).
  % [arXiv:nucl-th/0407067].
  %%CITATION = NUCL-TH 0407067;%%

\bibitem{fms_phot} 
  R.~J.~Fries, B.~M\"uller and D.~K.~Srivastava,
  Phys. \ Rev. \ Lett. {\bf 90}, 132301 (2003), and
  Phys. \ Rev. \ C {\bf 72}, 041902(R) (2005).

\bibitem{gale} 
  S.~Turbide, C.~Gale, and R.~J. Fries, 
  Phys. Rev. Lett. {\bf 96}, 032303 (2006).

\bibitem{inprep}
  R. Chatterjee, C. Gale, U. Heinz, and D.K. Srivastava, in preparation.

\bibitem{fn2}
A recent calculation by Hirano {\it et al.} \cite{Hirano_phi}, using a 
hybrid hydro+cascade approach where the evolution after hadronization is 
described by a realistic hadron cascade, predicts that the $\phi$ mesons 
have larger elliptic flow $v2(p_T)$ than shown in Fig.~\ref{F5}, similar 
to the blue dotted line in Fig.~\ref{F5} describing the hadronic dileptons. 
This is due to viscous effects in the hadronic phase which do not allow
the $\phi$ mesons to pick up quite as much radial flow as predicted
by the ideal fluid dynamical simulation presented here.

\bibitem{phenix_phi} S.~S.~Adler {\it et al.}, [PHENIX Collaboration],
Phys. Rev. C {\bf 72}, 014903 (2005).

\bibitem{Hirano_phi}
T. Hirano, U. Heinz, D. Kharzeev, R. Lacey, and Y. Nara, in preparation. 

\end{thebibliography}
\end{document}